      [1999/12/01 v1.4c Il Nuovo Cimento]
\documentclass{cimento}
\usepackage{graphicx}  
\usepackage{txfonts}
\usepackage{natbib}
\usepackage{longtable} 
\usepackage{epsfig} 
\newcommand\apj{ApJ}

\newcommand\aap{A\&A}
\newcommand\mnras{MNRAS}

\title{Panchromatic study of GRB 060124: from precursor to afterglow}

\author{P. Romano\from{ins:1}, 
	S.~Campana\from{ins:1},    
	G.~Chincarini\from{ins:1}\from{ins:2},
	J.~Cummings\from{ins:3}\from{ins:4},  
	G.~Cusumano\from{ins:5},    
	S.~T.~Holland\from{ins:3}\from{ins:6},
	V.~Mangano\from{ins:5},  
	T.~Mineo\from{ins:5},   
	K.~L.~Page\from{ins:7}, 
	V.~Pal'shin\from{ins:8},
	E.~Rol\from{ins:7},     
	T.~Sakamoto\from{ins:3}\from{ins:4},
	B.~Zhang\from{ins:9},
	R.~Aptekar\from{ins:8},	
	S.~Barbier\from{ins:3},  
	S.~Barthelmy\from{ins:3},
	A.~P.~Beardmore\from{ins:7},
	P.~Boyd\from{ins:3},       
	D.~N.~Burrows\from{ins:10},
	M.~Capalbi\from{ins:11},   
	E.~E.~Fenimore\from{ins:12},   
	D.~Frederiks\from{ins:8},  
	N.~Gehrels\from{ins:3},  
        P.~Giommi\from{ins:11},    
	M.~R.~Goad\from{ins:7},  
	O.~Godet\from{ins:7},     
	S.~Golenetskii\from{ins:8},
	D.~Guetta\from{ins:13},   
	J.~A.~Kennea\from{ins:10},
	V.~La~Parola\from{ins:5}, 
	D.~Malesani\from{ins:14}, 
	F.~Marshall\from{ins:3}, 
	A.~Moretti\from{ins:1},  
	J.~A.~Nousek\from{ins:10},
	P.~T.~O'Brien\from{ins:7}, 
	J.~P.~Osborne\from{ins:7}, 
	M.~Perri\from{ins:11},
	\atque
	G.~Tagliaferri\from{ins:1} 
}
\instlist{\inst{ins:1} INAF--Osservatorio Astronomico di Brera, Via E.\ Bianchi 46, I-23807 Merate (LC), Italy 	
	\inst{ins:2} Universit\`a{} degli Studi di Milano, Bicocca, Piazza delle Scienze 3, I-20126, Milano, Italy
	\inst{ins:3} NASA/Goddard Space Flight Center, Greenbelt, MD 20771, USA  
	\inst{ins:4} National Research Council, 2101 Constitution Avenue, NW TJ2114, Washington, DC 20418,  USA
  	\inst{ins:5} INAF--Istituto di Astrofisica Spaziale e Fisica Cosmica, Via U.\ La Malfa 153, I-90146 Palermo, Italy   
	\inst{ins:6} Universities Space Research Association, 10211 Wincopin Circle, S.\ 500, Columbia, MD 21044, USA
	\inst{ins:7} Department of Physics \& Astronomy, University of Leicester, LE1 7RH, UK
	\inst{ins:8} Ioffe Physico-Technical Institute, 26 Polytekhnicheskaya, St. Petersburg 194021, Russian Federation
  	\inst{ins:9} Department of Physics, University of Nevada, Las Vegas, NV 89154-4002, USA
  	\inst{ins:10} Dept.\ of Astronomy \& Astrophysics, Pennsylvania State University,  University Park, PA 16802, USA 
  	\inst{ins:11} ASI Science Data Center, Via G.\ Galilei, I-00044 Frascati (Roma), Italy 
	\inst{ins:12} Los Alamos National Laboratory, MS B244, NM 87545, USA
	\inst{ins:13} INAF--Osservatorio Astronomico di Roma, Via di Frascati 33, I-00040 Monteporzio Catone, Italy
	\inst{ins:14} International School for Advanced Studies (SISSA-ISAS), Via Beirut 2-4, I-34014 Trieste, Italy 
}
\PACSes{\PACSit{98.70.Rz}{$\gamma$-ray sources; $\gamma$-ray bursts}
\PACSit{01.30.Cc}{Conference proceedings}
}

\begin{document}
\shortauthor{P.\ Romano et al. }
\shorttitle{Prompt and afterglow emission of GRB~060124}
\maketitle

\begin{abstract}
We present observations of GRB~060124, the first event for which 
both the prompt and the afterglow emission could be observed simultaneously 
and in their entirety by the three Swift instruments and by   Konus-Wind. 
Thanks to these exceptional circumstances, the temporal and spectral properties 
of the prompt emission could be studied in the optical, X-ray and gamma-ray ranges (up to 2\,MeV).
While the X-ray emission (0.2--10~keV) clearly tracks the
gamma-ray burst, the optical component follows a different pattern,
likely indicating a different origin, possibly the onset of external
shocks.
The prompt GRB spectrum shows significant spectral evolution,
with both the peak energy and the spectral index varying. As observed in
several long GRBs, significant lags are measured between the hard- and
low-energy components, showing that this behaviour extends over 3
decades in energy.
The GRB peaks are also much broader at soft energies.
This is related to the temporal evolution of the spectrum, and can be
accounted for by the softening of the electron spectral index with
time.
The burst energy ($E_{\rm iso} \sim 5\times 10^{53}$ erg at $z=2.297$) 
and average peak energy ($E_{\rm p} \sim 300$~keV) 
make GRB\,060124 consistent with the Amati relation.
The X-ray afterglow is characterized by a 
decay which presents a break at $t_{\rm b} \sim 10^5$~s. 
\end{abstract}

\section{Introduction}

\begin{figure}[t]
\hspace{-0.8truecm}
\includegraphics[width=7cm,height=15cm,angle=270]{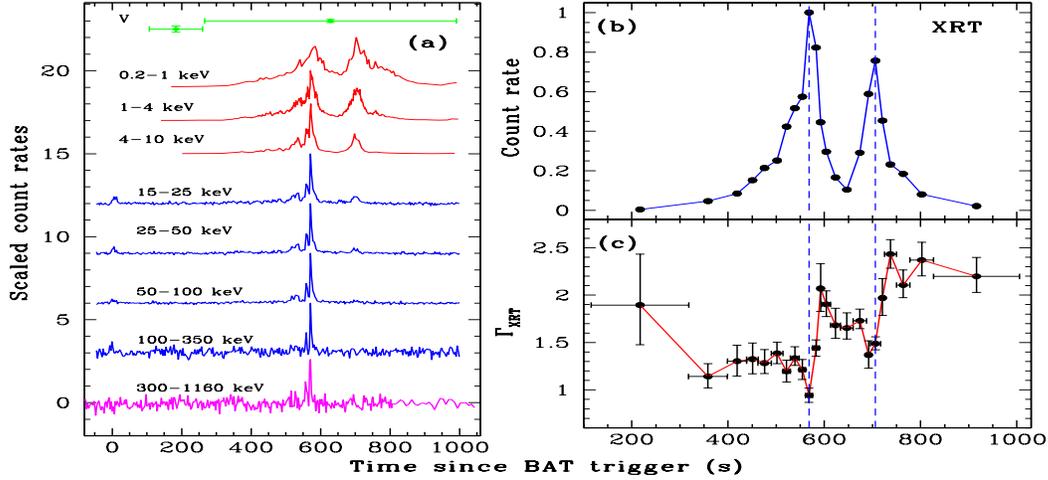} 
\caption{{\bf (a)} UVOT ($V$), XRT (0.2--10\,keV), BAT (15--350\,keV) and Konus 
		(300--1160\,keV) light curves. 
		The count rates were normalized to the peak of each light curve
		and offset vertically for clarity. \newline	
		Time-resolved spectroscopy of the prompt emission of GRB~060124 with XRT/WT data. 
		{\bf (b)}: Normalized XRT count rate. 
		{\bf (c)}: Time evolution of the photon index. 
		The vertical dashed lines mark the peak positions.
\label{venice:fig:alldata}}
\end{figure}

GRB~060124 is the first event for which the three Swift instruments 
have a clear detection of both the prompt and the afterglow emission. 
Indeed, Swift-BAT triggered on a precursor on 2006-01-24 at 15:54:52 UT, 
$\sim 570$\,s before the main burst peak. 
This allowed Swift to immediately repoint the NFIs
and acquire a pointing towards the burst $\sim 350$\,s before the burst occurred. 
The burst, which had a highly structured profile, comprises 
three major peaks following the precursor and had  one of longest 
total durations (even excluding the precursor) 
recorded by either BATSE or Swift (see Fig.~\ref{venice:fig:alldata}a). 
GRB~060124 also triggered Konus-Wind 
559.4\,s after the BAT trigger \cite{golenetskii2006:gcn4599}. 
The Konus light curve confirmed the presence of both the 
precursor and the three peaks of prompt emission. 
In this paper we summarize the observed properties of this burst. 
A fully detailed account of the observations and a more in-depth discussion 
of this exceptionally bright and long GRB 
(which is the very first case that the burst proper 
could be observed with an X-ray CCD with high spatial resolution imaging
down to 0.2\,keV) can be found in \cite{Romano2006:060124}. 

\section{The precursor/prompt phase}

\begin{figure}[t]
\hspace{-0.5truecm}
\includegraphics[width=7cm,height=6cm,angle=0]{romanop_fig2a.ps} 
\includegraphics[width=7cm,height=6cm,angle=0]{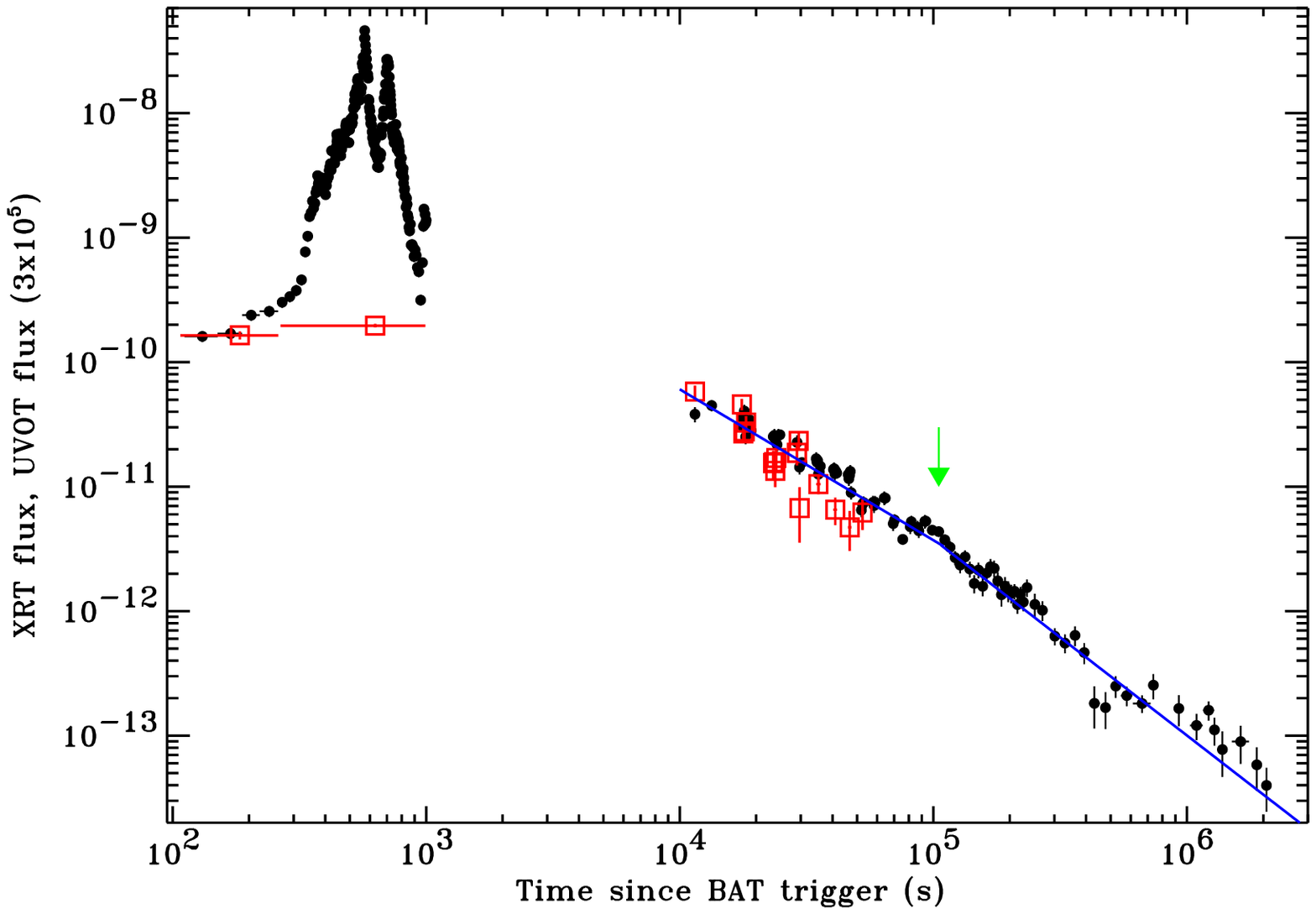} 
\caption{ {\bf (Left)}: Spectral energy distribution of GRB~060124 during the prompt phase.
		SEDs (1) and (2) are derived from simultaneous fits to XRT and BAT data fits, 
		while SEDs (3)--(5) from XRT, BAT and Konus fits.
{\bf (Right)}: XRT (filled circles) light curve in units of 
		erg s$^{-1}$ cm$^{-2}$ and UVOT (empty squares) flux density 
		(erg s$^{-1}$ cm$^{-2}$ \AA$^{-1}$) scaled to match the X-ray. 
		The solid line is the best fit model to the XRT 
		afterglow data. The arrow marks the X-ray break position. 
\label{venice:fig:alldata2}}
\end{figure}

	\begin{enumerate}
	\item For GRB~060124, we observed the longest interval ($\approx 500$\,s) 
		recorded between a precursor and the main GRB event 
		\citep{Lazzati2005:precursors}.
		The precursor is spectrally softer 
		(photon index $\Gamma_{\gamma,\,{\rm prec}}=1.8\pm0.2$) 
		than the following peaks ($\Gamma_{\gamma}=1.48\pm0.02$).
	\item The XRT, BAT and Konus light curves (0.2--2\,000\,keV, 
		see Fig.~\ref{venice:fig:alldata}a) 
		show the same overall structure consisting of three peaks.
		No time-resolved information is available from UVOT,
		but we can state that the optical behaviour during the prompt 
		emission was significantly different from the X-ray one.
		This may indicate the onset on a (weak) reverse shock.
	\item Based on the $T_{90}$ $\approx 300$\,s,  
		although the individual peak durations are comparable to 
		those of other GRBs, the total duration is among
		the longest recorded by either BATSE 	or Swift,  
		which may rule out precursor models in  which the fireball erupts from a massive 
		stellar envelope. 
	\item The peaks of the prompt emission shift with the energy band, with 
		the peaks observed in the harder bands preceding the ones observed 
		in the softer bands. This lag is as much as $\sim 10$\,s 
		(between 15--100\,keV and 0.2--1\,keV). 
		This has often been observed before \citep[e.g., ][]{Ford1995:softhard}; 
		our data show that that this behaviour extends over 3 decades in energy.

	\item Strong spectral evolution takes place during the prompt phase.
		{\it i)} We observe the peak energy $E_{\rm p}$ moving from higher energies to lower energies
		(Fig.~\ref{venice:fig:alldata2}a).
		{\it ii)} The spectral evolution follows a ``tracking'' behaviour
		(Fig.~\ref{venice:fig:alldata}c), with the
		spectrum being harder when the flux is higher.
		Furthermore, the photon index achieves a plateau following each peak 
		in the light curve, with each plateau becoming successively softer.  
		During this time interval, the photon index increases by about 1, 
		reaching a value consistent with the later afterglow spectrum 
		($\Gamma_{\rm X}=2.06\pm0.06$) by the end of the last peak.
		{\it iii)} The relative importance of the three main peaks 
		varies with the energy band (Fig.~\ref{venice:fig:alldata}a); the third peak in 
		the 0.1--1\,keV light curve is actually stronger than the second one, 
		as opposed to what is observed in all the other energy bands.
	\item The peaks are much broader in the softer bands (Fig.~\ref{venice:fig:alldata}a). 
		This is related to the temporal evolution of the spectrum, and can be
		accounted for by assuming that the electron spectral index softened with
		time.
	\item We derive a mean rest frame peak and isotropic energy 
		$E'_{\rm p}=636_{-129}^{+257}$\,keV 
		and $E_{\rm iso} = (4.2\pm0.5)\times 10^{53}$ erg. This burst is consistent
		with the Amati, the Yonetoku, and the Ghirlanda 
		\citep{Amati2002,Yonetoku2004,Ghirlanda2004} relations. 
	\end{enumerate}

\section{The afterglow phase}

	\begin{enumerate}
	\item The X-ray afterglow ($F(\nu,t) \propto t^{-\alpha} \nu^{-\beta}$) is modelled with a 
		broken power-law decay with indices $\alpha_1=1.21\pm0.04$, $\alpha_2=1.58\pm0.06$, 
		and a break at $t_{\rm b}=(1.05_{-0.14}^{+0.17})\times 10^{5}$\,s 
		(Fig.~\ref{venice:fig:alldata2}b). 
	\item The optical afterglow (data only available for $t<10^5$\,s) is modelled with 
		a power-law decay with index $\alpha_{BV} = 0.82\pm 0.06$ 
		(the optical data are only available for $t<10^5$\,s). 
		An optical break has been reported at a time consistent with the X-ray break
		\cite{Curran2007}, which makes this one of the few well-defined 
		achromatic breaks in the Swift sample \cite[see, ][]{Covino2007:swift_breaks}.
	\end{enumerate}

The early part of the afterglow ($t < t_{\rm b}$) can be explained well in
terms of the standard model. 
In this case, the optical and X-ray data constrain the cooling frequency 
$\nu_{\rm c}$ between the optical and X-ray bands. 
In fact, the X-ray and optical spectral indices 
$\beta_{\rm X} = 1.10 \pm 0.10$ and $\beta_{\rm opt} = 0.5 \pm 0.3$ differ by
$\approx 0.5$, and constrain $\nu_{\rm c} \sim 2 \times 10^{16}$~Hz. The
temporal behaviour is fully consistent with this interpretation: both the
optical and X-ray decay slopes are consistent with the model prediction:
$\alpha_{\rm X} \approx 3\beta_{\rm X}/2 - 1/2 = 1.15 \pm 0.10$ (valid for $\nu
> \nu_{\rm c}$) and $\alpha_{\rm opt} \approx 3\beta_{\rm opt}/2 = 0.75 \pm
0.40$ (valid for $\nu < \nu_{\rm c}$). The latter equation is valid only for a
homogeneous external medium (the wind model would require a much steeper
$\alpha_{\rm opt} \approx 1.3$). The electron distribution index is $p = 2\beta_{\rm
X} = 2.20 \pm 0.15$, which is in agreement with the shock theory.

The X-ray decay slope after the break is too shallow for the break being due to
a standard, sideways-expanding jet, in which case $\alpha_{\rm X} = p > 2$ would
be expected \citep{Rhoads1999:beaming}. 
The break magnitude $\Delta\alpha_{\rm X} = 0.39
\pm 0.08$ may be consistent with a jet which propagates in a wind environment
suffering no sideways expansion (for which $\Delta\alpha = 0.5$). However, the
wind model is ruled out by the early-time optical data. A solution which
satisfies all the available constraints is a structured jet having a
homogeneous core surrounded by power-law wings with energy profile ${\rm
d}E/{\rm d}\Omega \propto \vartheta^{-q}$  \citep{Panaitescu2005:jets}. 
The break magnitude is dependent upon the index $q$ (and on the location of
$\nu_{\rm c}$). With the observed values, we infer $q \approx 0.85$. The break
magnitude below the cooling frequency (i.e. in the optical) is slightly
different: $\Delta\alpha_{\rm opt} = 0.47$. 
This difference, however, is quite small.

\acknowledgments
This work was supported at INAF by ASI grant I/R/039/04.


\end{document}